\documentclass[twocolumn,pra,amssymb,amsmath,natbib,showpacs,%
floatfix,superscriptaddress]{revtex4}%
\usepackage[dvips]{graphicx}
\usepackage{color}
\usepackage[final]{showkeys}

\newcommand{\Eref}[1]{{Eq.~(\ref{#1})}}
\newcommand{\Fref}[1]{{Fig.~\ref{#1}}}
\newcommand{\etal}{{\it et al.}}

\begin{document}

\title{Fractional exclusion statistics and the Random Matrix Boson Ensemble}%

\author{Sa\'ul Hern\'andez-Quiroz}%
%\email{saul@cicc.unam.mx}%
\affiliation{%
  Instituto de Ciencias F\'{\i}sicas, Universidad Nacional Aut\'onoma
  de M\'exico (UNAM), 62210-Cuernavaca, Mor.,
  M\'exico}%
\affiliation{%
  Facultad de Ciencias, Universidad Aut\'onoma del Estado de Morelos
  (UAEM), 62209-Cuernavaca, Mor., M\'exico }%
\author{Manuel Beltr\'an}
\affiliation{%
  Instituto de F\'isica, Universidad Nacional Aut\'onoma de M\'exico (UNAM),
  M\'exico DF, M\'exico}%
\author{Luis Benet}%
%\email{benet@fis.unam.mx}%
\affiliation{%
  Instituto de Ciencias F\'{\i}sicas, Universidad Nacional Aut\'onoma
  de M\'exico (UNAM), 62210-Cuernavaca, Mor.,
  M\'exico}%
\author{Jorge Flores}%
\affiliation{%
  Instituto de F\'isica, Universidad Nacional Aut\'onoma de M\'exico (UNAM),
  M\'exico DF, M\'exico}%
\author{Germinal Cocho}%
\affiliation{%
  Instituto de F\'isica, Universidad Nacional Aut\'onoma de M\'exico (UNAM),
  M\'exico DF, M\'exico}%

\date{\today}

\begin{abstract}
   The $k$--body Gaussian Embedded Ensemble of Random Matrices is considered 
   for $N$ bosons distributed on two single-particle levels. When $k = N$, the 
   ensemble is equivalent to the Gaussian Orthogonal Ensemble (GOE), and when 
   $k = 2$ it corresponds to the Two-body Random Ensemble (TBRE) for bosons. It 
   is shown that the energy spectrum leads to a rank function which is of the form
   of a discrete generalized beta distribution. The same distribution is obtained 
   assuming $N$ non-interacting quasiparticles that obey the fractional exclusion 
   statistics introduced by Haldane two decades ago.
\end{abstract}

%  05.30.-d Quantum statistical mechanics
%  05.30.Pr Fractional statistics systems
%  05.30.Jp  Boson systems

\maketitle

\section{Introduction}
Fractional exclusion statistics was introduced~\cite{Haldane1991} as a 
generalization of the Pauli exclusion statistics. The particles obeying such 
statistics are elementary excitations that can only exist in the interior of the 
whole system of interacting particles. It has also been shown that the 
excitations in exactly soluble many-body problems in one dimension, e.g. 
the Colagero-Sutherland model~\cite{Calogero1969,Sutherland1971}, the 
quasi-particles in the Luttinger model~\cite{Haldane1981} and the $\delta$ 
function gas~\cite{Murty1994}, obey this statistics. Excitations in the 
Haldane-Shastry spin model follow FES also~\cite{Haldane1988}. In fact, it has 
rigorously been proved that quasi-particles with non-trivial exclusion statistics 
exist in a class of models that are solved by the Bethe ansatz~\cite{Isakov1994}. 
Applications to other condensed matter systems have also been 
considered~\cite{Khare2005}.

It has been shown~\cite{Iguchi2000} that a quantum liquid of particles 
interacting via a long-range two-body potential, where the particles are 
supposed to be either bosons or fermions, exhibits the nature of a quantum 
liquid with FES. The expectation that FES in arbitrary dimensions continuously 
interpolates between Fermi and Bose liquids holds only for one-dimensional 
systems. Recently~\cite{Anghel2008}, it has been shown that FES is a consequence 
of the interactions between the particles of the system and is due to the change 
from the description in terms of free-particle energies to the description in terms of 
quasiparticle energies.

On the other hand, several models in which two-level systems are occupied by 
bosons have been useful to understand different quantum many-body systems and 
can be treated numerically by direct diagonalization. These models have as a common 
feature that the lower level has a scalar boson, but differ according to the multipolarity 
$L$ of the second level. For example, the Lipkin-Meshkov-Glick model~\cite{Lip65} in 
the Schwinger representation has a scalar $L=0$ boson in the upper level, as is the 
case in the two-mode Bose-Hubbard model for Bose-Einstein condensates confined in 
two-well potentials~\cite{Milburn1997}. A dipolar ($L=1$) boson leads to the vibron 
model of quantum chemistry and a quadrupole ($L=2$) boson corresponds to the 
interacting boson model~\cite{Iachello1987}.

Here we shall study a rather simple, yet general, two-level system with $N$ spin-less 
bosons interacting through a $k$--body random potential; this is called the $k$--body 
Gaussian Embedded Ensemble for bosons~\cite{Asaga2001,BW2003}. More 
specifically, we deal with the cases $k=2$ and $k=N$. The former case corresponds 
to the most common physical situation with two-body interactions. The latter 
corresponds to the canonical ensembles of Random Matrix Theory 
(RMT)~\cite{Brody81}, which is important in its own since RMT predicts accurately 
the universal statistical properties of the fluctuations of the spectra of a large variety 
of complicated physical systems~\cite{GMGW1998}.

In this paper we use the $k$--body boson random ensemble to calculate the 
integrated level density and show that, in all cases, beta-like rank distributions are 
obtained. Such beta-like distributions have been observed to hold in other complex 
systems in biology, social sciences and arts~\cite{MartinezMekler}. We also obtain 
here this distribution using quasiparticles that obey FES. This implies a generalization 
of two-level systems, since the second level must have a degeneration which is no 
longer an integer number, as can occur with Haldane particles~\cite{Haldane1991} 
as well as with Regge poles~\cite{Regge1959}.

\section{Two-level boson systems with random interactions}
\label{bosons}
The $k$--body Gaussian Embedded Ensemble for bosons is defined as 
follows~\cite{Asaga2001,BW2003,HQB2010}. We denote by $k$ the rank of the 
interaction and by $N$ the number of spin-less (scalar) bosons so $k \leq N$. The 
$N$ bosons are distributed in two single-particle levels; for simplicity, we assume 
that these levels are degenerate. We introduced the boson creation and annihilation 
operators $a_j^\dagger$ and $a_j$ ($j = 1, 2$), which satisfy the usual commutation 
relations for bosons. The normalized $N$-boson states are denoted by 
$|\mu_n^{(N)}\rangle = ({\cal A}_{n,N})^{-1} 
(\hat a_1^\dagger)^n (\hat a_2^\dagger)^{N-n}|0\rangle$, where 
${\cal A}_{n,N} = [n! (N-n)!]^{1/2}$ is a normalization constant and $|0\rangle$ is 
the vacuum state. Clearly, the dimension of the Hilbert space is $W_{\rm B}=N+1$ 
(see \Eref{Wg} below with $g=0$). Then, in second-quantized form, the most 
general Hamiltonian $\hat H_k^{(\beta)}$ involving $k$--body interactions is given by 
\begin{equation}
  \label{Hkbosons}
  {\hat H_k^{(\beta)}} = \sum_{n,m=0}^k \, v_{n,m}^{(\beta)} \,
  \frac{ (\hat a_1^\dagger )^{n} ( \hat a_2^\dagger )^{k-n} 
    (\hat a_1 )^{m} ( \hat a_2 )^{k-m} } {{\cal A}_{n,k}\, {\cal A}_{m,k}}  \ .
\end{equation}

\begin{figure}
  \includegraphics[width=8.5cm]{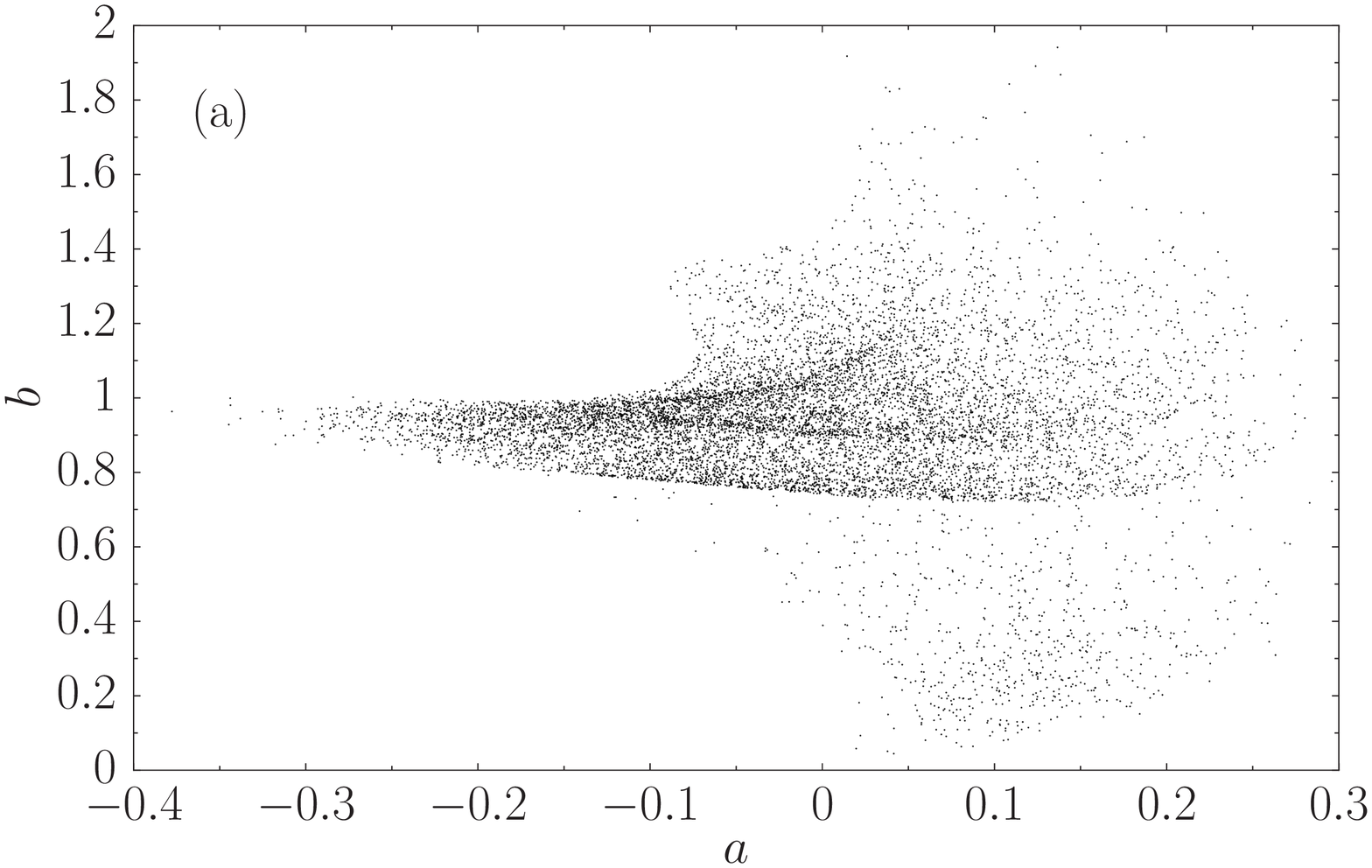}
  \includegraphics[width=8.5cm]{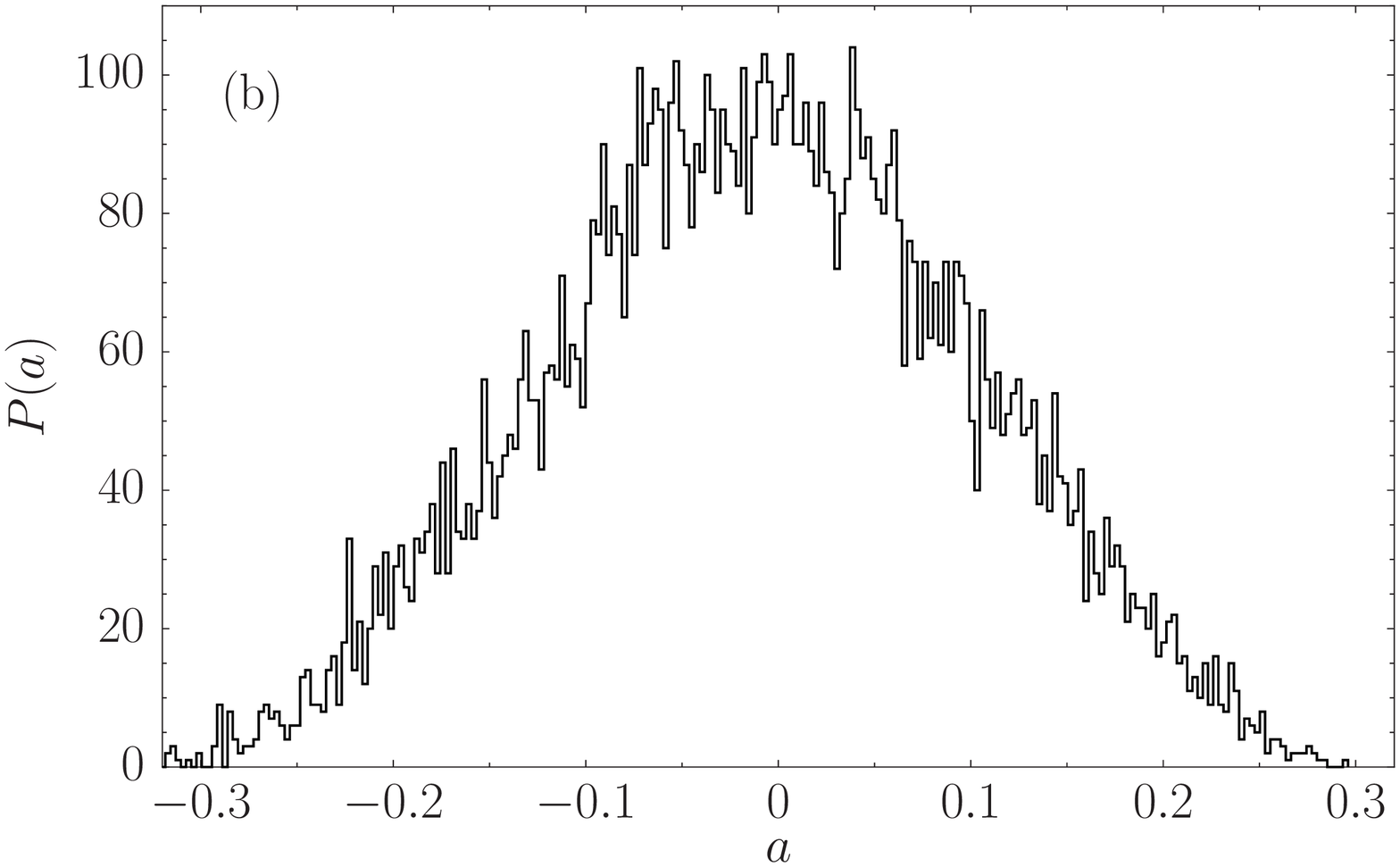}
  \includegraphics[width=8.5cm]{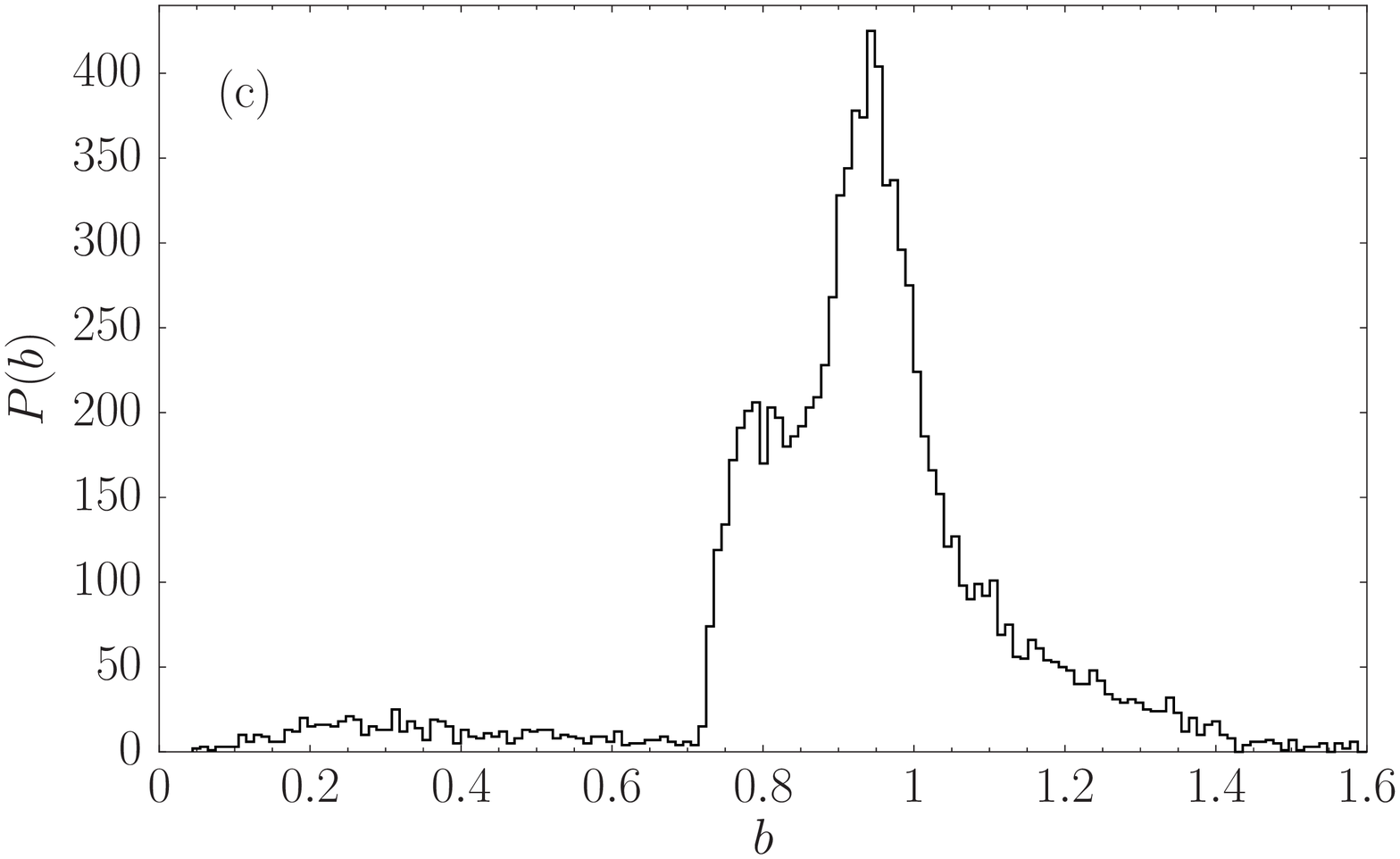}
  \includegraphics[width=8.5cm]{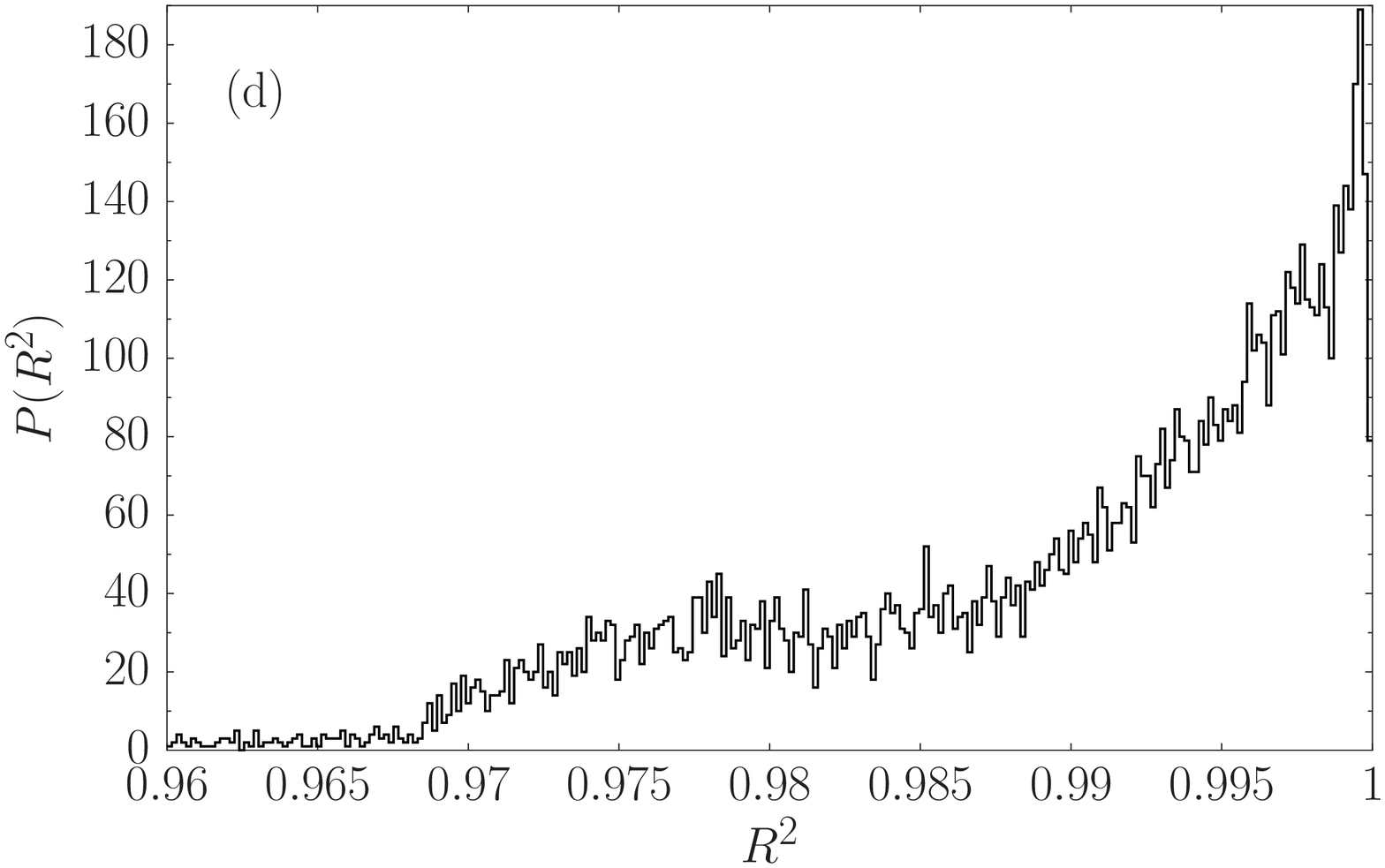}
  \caption{ (a)~Scatter plot of the fitting parameters $a$ and $b$ of
    \Eref{DGBD} computed for each realization of the interacting
    two-level bosons ensemble for $k=2$ and $N=2000$. The frequency
    distributions of $a$, $b$, and $R^2$ are given in figures (b),
    (c), and (d), respectively. }%
  \label{fig1}%
\end{figure}

Stochasticity is built into the Hamiltonian $\hat H_k^{(\beta)}$ at
the level of the $k$--body matrix elements $v_{n,m}^{(\beta)}$. These
matrix elements are assumed to be Gaussian distributed independent
random variables, with zero mean and constant variance $v_0^2=1$. As
in the case of RMT~\cite{GMGW1998}, Dyson's parameter $\beta$
distinguishes the ensembles according to its symmetry properties:
$\beta=1$ corresponds to the case where time--reversal symmetry holds,
and $\beta=2$ holds when time--reversal invariance is broken.
The $k$--body interaction matrix $v^{(\beta)}$ is thus a
member of the Gaussian Orthogonal Ensemble (GOE) for $\beta=1$ or of the
Gaussian Unitary Ensemble (GUE) for $\beta=2$~\cite{GMGW1998}. 
 
\begin{figure}
  \includegraphics[width=8.5cm]{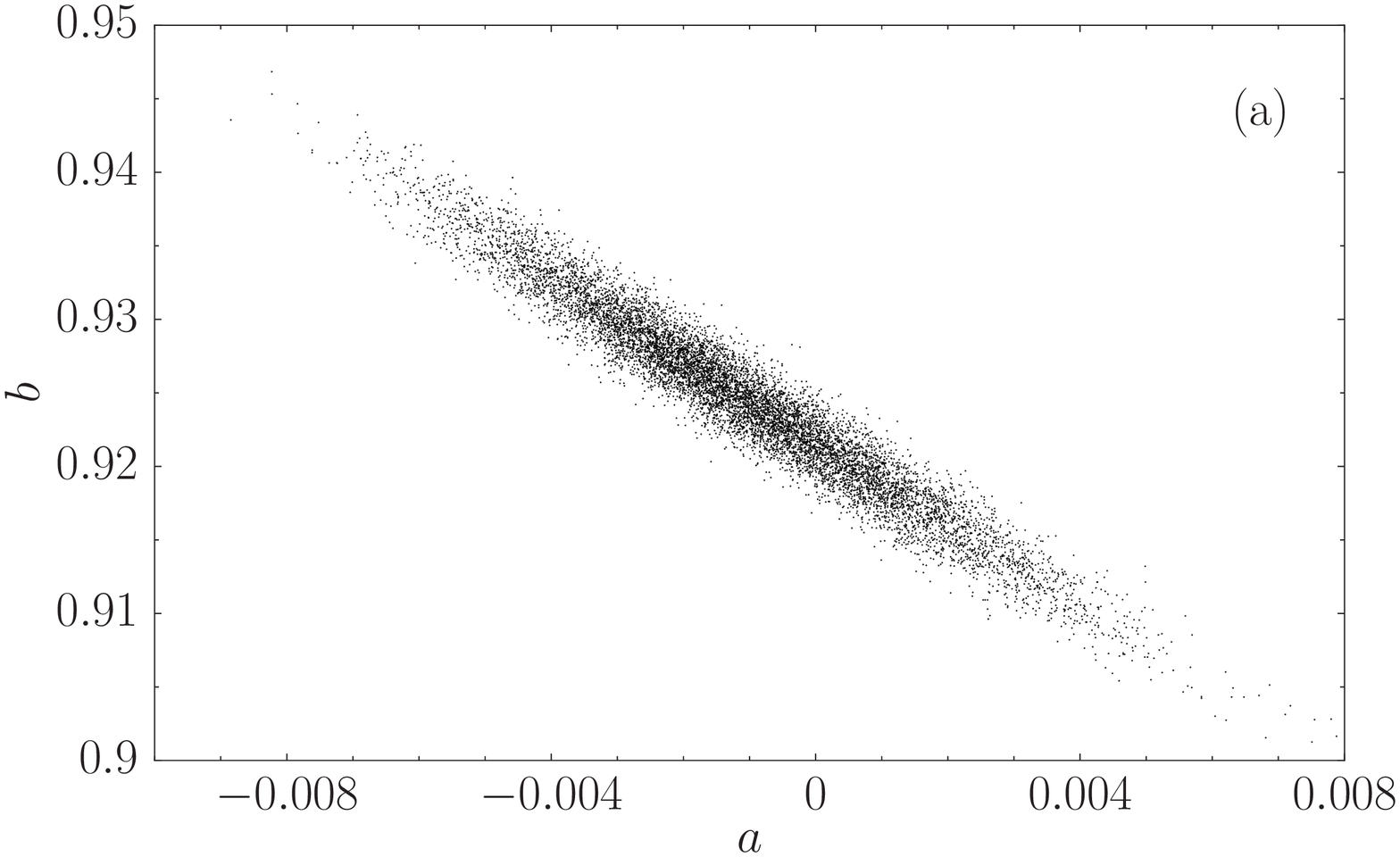}
  \includegraphics[width=8.5cm]{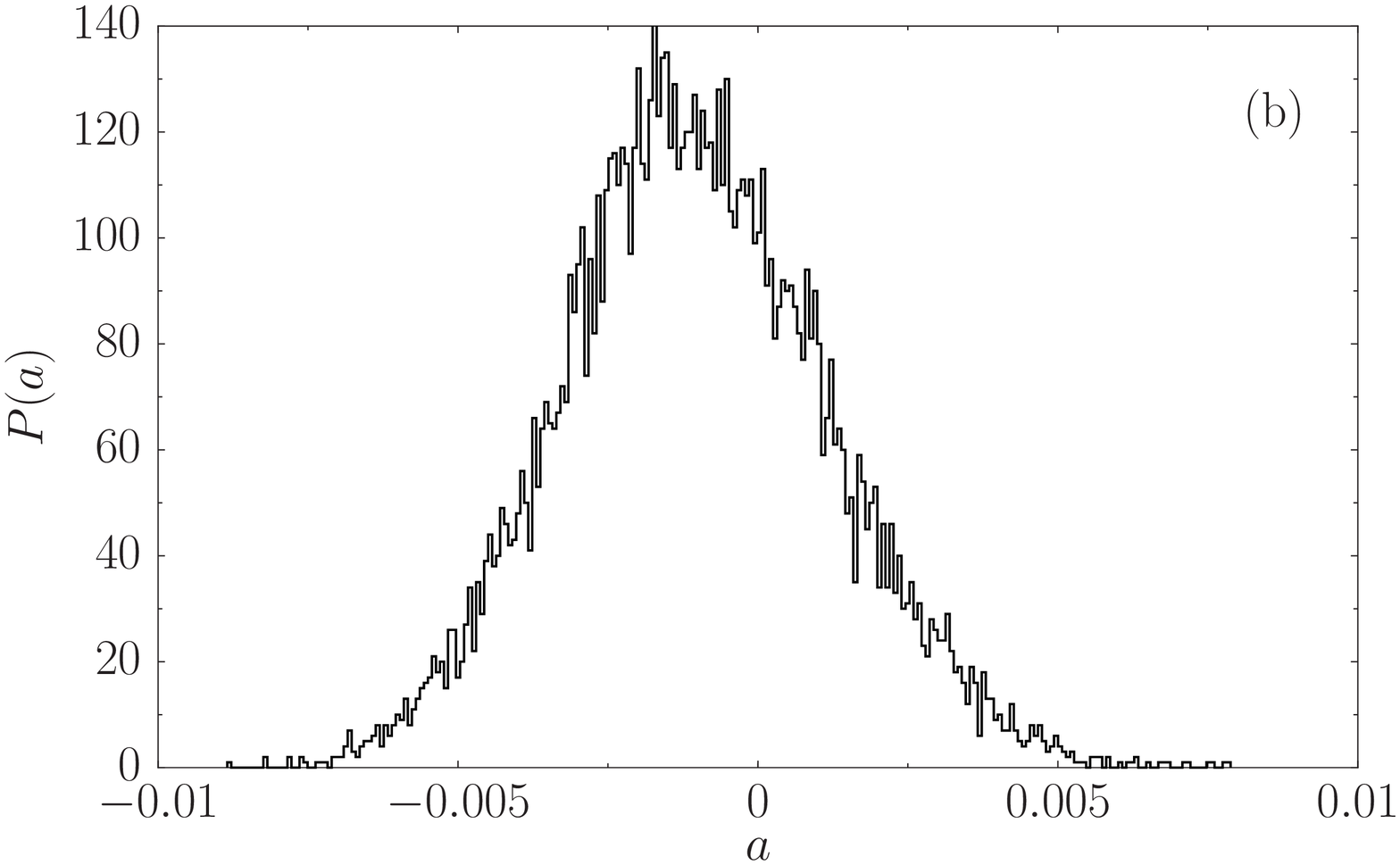}
  \includegraphics[width=8.5cm]{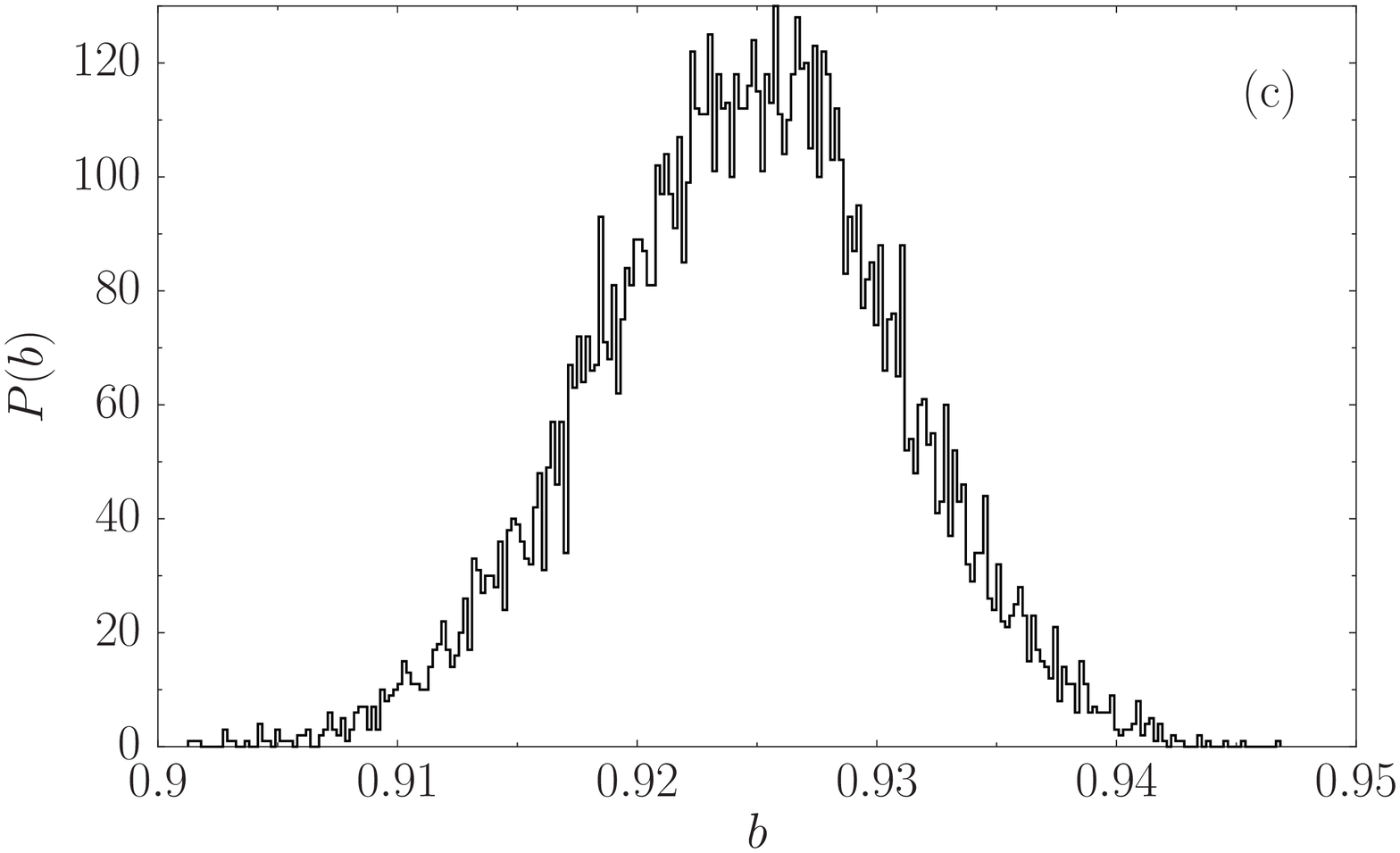}
  \includegraphics[width=8.5cm]{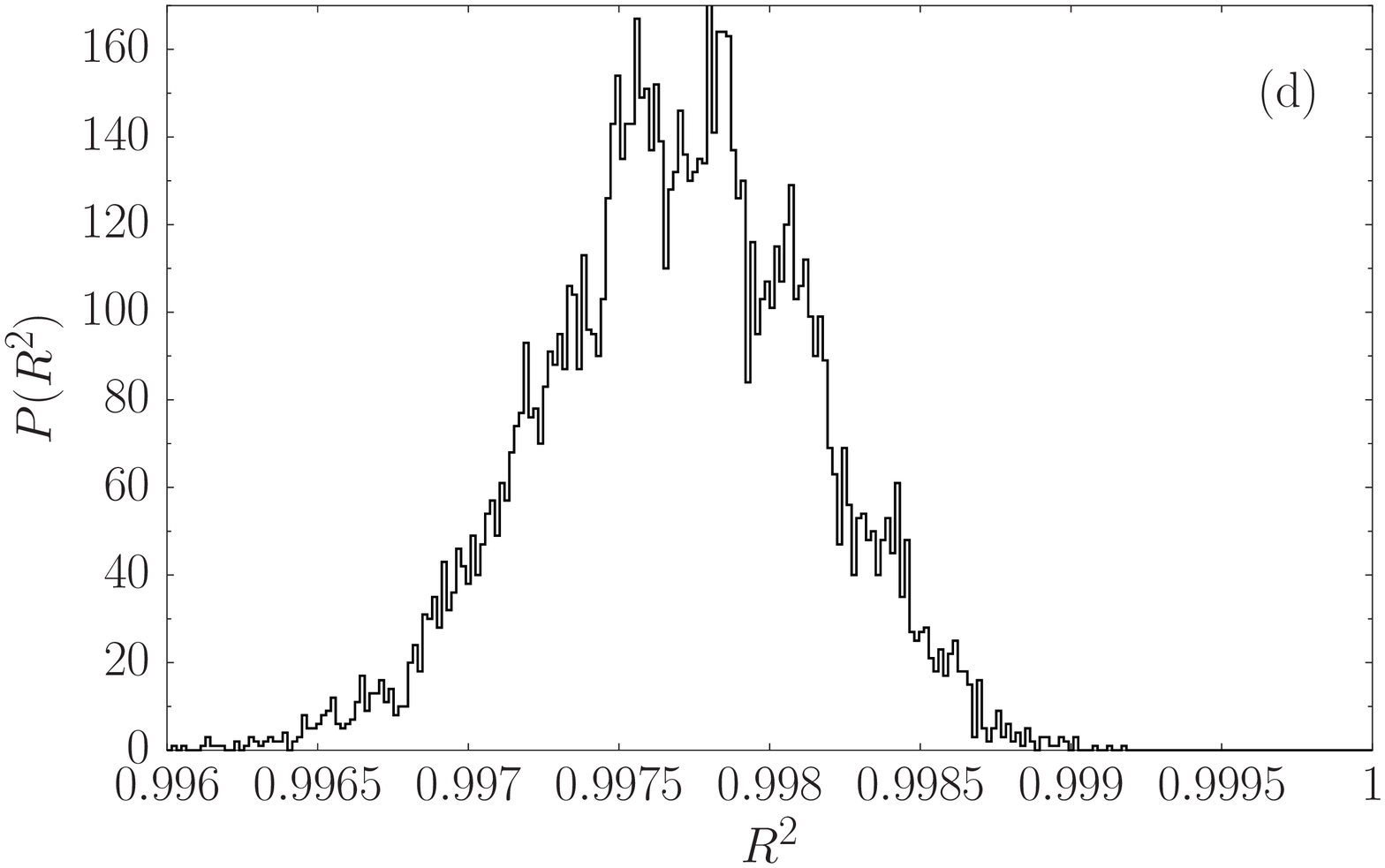}
  \caption{
    Same as \Fref{fig1} for $k=N$.}%
  \label{fig2}%
\end{figure}

We note that the number of independent matrix elements of the
$k$--body matrix is equal to
$K_\beta(k)=\beta(k+1)(k+1+\delta_{\beta,1})/2$. Therefore, for $k\ll
N$, which is physically the relevant case, the matrix elements of
$\hat H_k^{(\beta)}$ are correlated, i.e., the number of independent
random variables is much smaller than the number of independent matrix
elements. In addition the matrix is sparse since many matrix elements
are identically zero. The combinatorial factors ${\cal A}_{n,k}$ and
${\cal A}_{m,k}$ are introduced in the definition of the ensemble
\Eref{Hkbosons} in order to have for $k=N$ perfect identification,
i.e., a one-to-one correspondence, with the canonical ensembles of
RMT~\cite{Asaga2001,BW2003}.

The numerical results presented below are related to the integrated
level density, also referred to as the staircase function, of
individual realizations of the ensemble~(\ref{Hkbosons}).
We take into account a large number of such realizations to achieve small 
statistical errors. For each member of the ensemble we construct the staircase 
function directly from the diagonalization of the Hamiltonian. The staircase 
function is defined as $\mathfrak{N}(E) =\Theta (E-\hat H_k^{(\beta)})
=\sum_i\,\Theta(E-E_i)$, where $E_i$ are the many-body eigenvalues of
the specific realization of the Hamiltonian; $\Theta(x)$ is the
Heaviside step function. By definition, the staircase function is a
monotonic non-decreasing function of the energy. For later convenience
we shall use the rank function $r(E)= N+1-\mathfrak{N}(E)$. We invert
it numerically and express the energy $E$ in terms of the rank $r$;
the function $E(r)$ is therefore a monotonic decreasing function of
the rank $r$. For each member of the ensemble, the rank distribution
$E(r)$ is fitted with the two-parameter
distribution~\cite{MartinezMekler}:%
\begin{equation}
  \label{DGBD}
  f(r) = A\frac{(r_{\rm max}+1-r)^b}{r^a}.
\end{equation}
Here, $A$ is the normalization constant of the distribution, $r_{\rm
  max}=N+1$ is the maximum value attained by the rank $r$, and the
constants $a$ and $b$ are two fitting exponents. We refer to $f(r)$ as
the discrete generalized beta distribution (DGBD).

We present now numerical results on the fitting exponents $a$ and $b$
of Eq.~\ref{DGBD}. These were obtained by considering $10000$ members
of the $k$-body Gaussian Embedded Ensemble for two-level bosons, where
we fixed $N=2000$ bosons and $\beta=1$; similar results were obtained
for $\beta=2$. The best fit was determined by a log-log multiple
linear regression; the quality of the fit is measured by the square
correlation coefficient $R^2$, which must be close to $1$ for a
reliable fit. We focus on the cases $k=2$ and $k=N$ since these are
physically the most relevant. In \Fref{fig1} we present a scatter plot
of the fitting parameters obtained for each individual member of the
ensemble for $k=2$ and the frequency distributions of $a$, $b$, and
$R^2$; \Fref{fig2} displays the corresponding results for
$k=N$. Interestingly, these figures show that the distributions for
$a$ and $b$ are much broader for $k=2$ than for $k=N$; the results on
$R^2$ simply give confidence to the fitting procedure.

From the numerical results, we compute the average values of the
fitting parameters. For $k=2$ we have $\bar{a}_{k=2}=-6.43\times
10^{-3}\pm 0.1094$ and $\bar{b}_{k=2}= 0.922\pm 0.209$, where the
assigned error corresponds to the standard deviation of the sample.
Likewise, for $k=N$ we obtain $\bar{a}_{k=N}=-8.92\times 10^{-4}\pm
2.15\times 10^{-3}$ and $\bar{b}_{k=N}=0.924\pm 6.17\times 10^{-3}$.
In both cases, $\bar{a}$ is slightly negative and very close to zero, and
$\bar{b} < 1$. We shall show in Section~\ref{RankDists} that the fact
that $\bar a \simeq 0$ and $\bar b \lesssim 1$ indicates that interactions mimic
FES, i.e., interacting particles are equivalent to non-interacting
Haldane quasi-particles with respect to the rank distribution.

\section{Rank distributions for fractional exclusion statistics}
\label{RankDists}

The combinatorial formula for the number of many--body states $W_g$ of
$N$ identical particles following FES and occupying a group of $p$
states is given by (cf. page 147 of~\cite{Muirhead1965})
\begin{equation}
  \label{Wg1}
  W_g = \frac{[p+(N-1)(1-g)]!}{N!\,[p-gN-(1-g)]!}.
\end{equation}
This expression reproduces for $g=0$ the well-known expression for
bosons, $W_{\rm B} = [p+N-1]!/[N!\,(p-1)!]$, and for $g=1$ that of
fermions $W_{\rm F} = p!/[N!\,(p-N)!]$. Note that \Eref{Wg1} can be
rewritten as
\begin{equation}
  \label{Wg}
  W_g = \frac{[p'+N-1]!}{N!\,[p'-1]!},
\end{equation}
with
\begin{equation}
  \label{pprime}
  p'=p-g(N-1).
\end{equation}

One has therefore $N$ identical bosons occupying an {\it effective}
group of $p'$ single-particle states. Note that $p'$ could be formally
negative. This gives the possibility of an interesting connection with
the famous Regge trajectories used in high-energy physics~\cite{Muirhead1965}.

In order to compare the rank distribution of $N$ particles obeying FES
with those obtained in Section~\ref{bosons} for bosons, we shall group
the $p'$ single-particle levels filled by $N$ Haldane particles into
two levels with energies $\epsilon_0=0$ and $\epsilon_1>\epsilon_0$,
with $p'=p_0'+p_1'$.  If $p_0'=1$, the number of (non-interacting Haldane
particles) microstates with energy $E=n_1\epsilon_1$ is given by%
\begin{equation}
  \omega(n1=E/\epsilon_1) = \frac{(n_1+p'_1-1)!}{n_1!\,(p'_1-1)!}\, .
\end{equation}
The rank distribution with respect to the energy is then
\begin{equation}
  r(E)=\sum_{\nu=E/\epsilon_1}^N \omega(\nu),
\end{equation}
which for $p'_0=1$ yields
\begin{eqnarray}
  \label{ OJO }
  r(E) & =& \sum_{\nu=0}^N \omega(\nu)-
  \sum_{\nu=0}^{E/\epsilon_1} \omega(\nu) \nonumber\\
  & = &
  \frac{(N+p_1)!}{N!\, p_1!}-\frac{(E-\epsilon_1+p_1)!}{(E-\epsilon_1-1)!\,p_1!}\, .
\end{eqnarray}
Taking the continuum limit of $E$ and using Stirling formula, we obtain
\begin{equation}
  E \simeq (N^{p'_1}-p'_1! r )^{1/p'_1}\, .
\end{equation}

For particles obeying FES the effective number of states $p'_1$ can be
positive or negative, as we have remarked in connection with
\Eref{pprime}. If $p'_1>0$ and $N$ is large, we have
\begin{equation}
  \label{positivecase}
  \frac{E}{(p'_1!)^{1/p'_1}} = ({\cal N}-r)^{1/p'_1}\, ,
\end{equation}
with ${\cal N}=N^{p'_1} /p'_1 !$, which is a discrete generalized beta
distribution  with $a=0$ and
$b=1/p'_1$, as seen from direct comparison with $f(r)$
in~\Eref{DGBD}. On the other hand, if $p'_1<0$ and $N$ is large, we
have a rank distribution of the form
\begin{equation}
[(-p'_1)!r]^{-1/p'_1} \, ,
\end{equation}
which corresponds to the values $a=1/|p'_1|$ and $b=0$ of the DGBD. In
order to have $(-p'_1)!$ real, $p'_1$ must satisfy $-2s-1<p'_1<-2 s$,
with $s$ a positive integer.

\section{Discussion}
\label{Discussion}
We have given numerical evidence to show that the $k$-body embedded 
random matrix ensemble for bosons leads to an integrated level density 
which corresponds to a discrete generalized beta distribution. We have 
defined the ensemble for $N$ bosons interacting through a $k$-body force
distributed in two single-particle levels. When $k = N$ the GOE is obtained
by construction, and the two-body random ensemble~\cite{French1970} for 
bosons corresponds to $k = 2$, a more physical case. 

The discrete generalized beta distribution fits rather well the statistical 
regularities in many instances: in particular, for the rank-citation profile of 
scientists~\cite{Peterson}. This distribution is present in systems whose output 
depends on the restricted difference of statistical variables~\cite{Beltran}. Examples 
of applicability are neuron dynamics (excitation and inhibition) and ecological 
community evolution (birth and death).

We then considered a set of independent Haldane quasiparticles, which obey 
fractional exclusion statistics, distributed in the same two-level system. It is shown 
that DGBD is also obtained. With respect to this distribution, at least, the above 
results indicate that the interaction among the particles makes the system 
equivalent to a set of quasiparticles obeying FES.

\begin{acknowledgments} 
We acknowledge financial support from the projects IN-114310 (DGAPA-UNAM), 
57334-F (CONACyT). LB acknowledges CONACyT for the sabbatical grant 144684-ES1.
{\bf XXX}.
\end{acknowledgments}

\end{document}